\documentclass{pasa}%

\usepackage{graphicx}
\usepackage{txfonts}

\usepackage{microtype}

\title[The linear polarisation structure of Centaurus A]{The extraordinary linear polarisation structure of the southern Centaurus A lobe revealed by ASKAP}

\author[Craig Anderson et al.]{Craig S. Anderson$^1$\thanks{Correspondence: craig.anderson@csiro.au}, George Heald$^1$, Shane P. O'Sullivan$^2$, John D. Bunton$^3$, Ettore Carretti$^4$, Aaron P. Chippendale$^3$, Jordan D. Collier$^{5,6}$, Jamie S. Farnes$^7$, Bryan M. Gaensler$^8$, Lisa Harvey-Smith$^{9,10}$, B\"{a}rbel S. Koribalski$^3$, Tom L. Landecker$^{11}$, Emil Lenc$^3$, Naomi M. McClure-Griffiths$^{12}$, Daniel Mitchell$^3$, Lawrence Rudnick$^{13}$, Jennifer West$^8$
\affil{$^1$ \quad CSIRO Astronomy and Space Science, Kensington, Perth 6151, Australia}
\affil{$^2$ \quad Hamburger Sternwarte, Universit\"{a}t Hamburg, Gojenbergsweg 112, 21029 Hamburg, Germany}
\affil{$^3$ \quad CSIRO Astronomy and Space Science, PO Box 76, Epping NSW 1710, Australia}
\affil{$^4$ \quad INAF-Istituto di Radioastronomia, Via Gobetti 101, Bologna, Italy}
\affil{$^5$ \quad School of Computing, Engineering and Mathematics, Western Sydney University, Locked Bag 1797, Penrith, NSW 2751, Australia}
\affil{$^6$ \quad The Inter-University Institute for Data Intensive Astronomy (IDIA), Department of Astronomy, University of Cape Town, Rondebosch, 7701, South Africa}
\affil{$^7$ \quad Oxford e-Research Centre (OeRC), Department of Engineering Science, University of Oxford, Oxford, OX1 3QG, UK}
\affil{$^8$ \quad Dunlap Institute for Astronomy and Astrophysics, 50 St. George Street, Toronto, ON M5S 3H4, Canada}
\affil{$^9$ \quad School of Physics, The University of New South Wales, Kensington 2033, New South Wales, Australia}
\affil{$^{10}$ \quad School of Computing, Engineering and Mathematics, Western Sydney University, Locked Bay 1797, Penrith NSW 2751, Australia}
\affil{$^{11}$ \quad Dominion Radio Astrophysical Observatory, Herzberg Astronomy, National Research Council Canada, PO Box 248 Penticton B.C. Canada V2A 6K3}
\affil{$^{12}$ \quad Research School of Astronomy and Astrophysics, Australian National University, Canberra ACT 2611, Australia}
\affil{$^{13}$ \quad Minnesota Institute for Astrophysics, University of Minnesota}

}%

\jid{PASA}
\doi{10.1017/pas.\the\year.xxx}
\jyear{\the\year}
\usepackage{aas_macros}
\usepackage{hyperref} 
\hypersetup{colorlinks,citecolor=blue,linkcolor=blue,urlcolor=blue}

\hypersetup{draft}

\begin{document}

\begin{frontmatter}
\maketitle

\begin{abstract}
We present observations of linear polarisation in the southern radio lobe of Centaurus A, conducted during commissioning of the Australian Square Kilometre Array Pathfinder (ASKAP) telescope. We used 16 antennas to observe a 30 square degree region in a single 12 hour pointing over a 240 MHz band centred on 913 MHz. Our observations achieve an angular resolution of $26\times33$ arcseconds (480 parsecs), a maximum recoverable angular scale of 30 arcminutes, and a full-band sensitivity of 85 $\muup$Jy beam$^{-1}$. The resulting maps of polarisation and Faraday rotation are amongst the most detailed ever made for radio lobes, with of order 10$^5$ resolution elements covering the source. We describe several as-yet unreported observational features of the lobe, including its detailed peak Faraday depth structure, and intricate networks of depolarised filaments. These results demonstrate the exciting capabilities of ASKAP for widefield radio polarimetry.

\end{abstract}

\end{frontmatter}

\section{Introduction}

The outer lobes of Centaurus A (hereafter, Cen A) are unique in our sky. They are generated by our closest active galactic nucleus, and are found 10 times closer to us than expected on the basis of their type and luminosity \citep{MS2007}. Thus, they represent a peerless target for studying the processes occurring in and around FR-I-type radio galaxies.

Cen A's radio emission is complex and structured on all scales at which it has been observed (e.g. see \citealp{Junkes1993,Israel1998,Morganti1999,Ojha2010,Feain2011,Muller2014,McKinley2018}). It is precisely this structure that has given us important insights into as varied a set of astrophysical processes as jet launching and collimation (e.g. \citealp{Ojha2010,Muller2014}), jet-environment interactions (e.g. \citealp{Morganti2010} and refs. therein, also \citealp{TL2009,NEO2015,Wykes2015,McKinley2018}), MHD instabilities \citep{Wykes2014}, astrophysical shocks and particle acceleration processes (e.g.\citealp{Croston2009,Crockett2012,Wykes2015b}), and more. 

Many of these studies were made possible by sub-arcminute resolution imaging of the lobes' total intensity radio emission. It is therefore unfortunate that polarimetric imaging of the outer lobes has not kept pace, with the current highest resolution maps only resolving structures 4.3 arcminutes (4.7 kpc) across at 4.8 GHz \citep{Junkes1993}, and 14 arcminutes (15 kpc) across at 1.4 GHz \citep{OSullivan2013}. There are compelling reasons to obtain polarimetric data at higher resolution: \citet{Feain2009} demonstrate that on scales from 0--20 arcminutes, the lobes contribute to the Faraday rotation of background radio sources, prompting questions such as where the Faraday active plasma is located in the system, how it came to be there, and what it might tell us about the dynamics and physical history of the lobes. \citet{OSullivan2013} detect $\sim10^{10}$ M$_\odot$ of this plasma mixed throughout the lobes, with important implications for the questions just posed. Nevertheless, without higher resolution data, it is difficult to state with confidence how this material is distributed in the system.

The large size of Cen A makes it expensive to observe: To obtain the multi-frequency, high spatial resolution data required to properly interpret its linearly polarised emission would traditionally have required several thousand hours of observations (see \citealp{Feain2011}). This has recently changed with innovative new aperture synthesis arrays such as ASKAP (Section \ref{sec-askap}), characterised by wide correlated bandwidths and exceptional survey speeds. With ASKAP, we have been able to commence a detailed study of the Cen A lobes in full polarisation at sub-arcminute resolution with just 12 hours of on-sky data. Our eventual aim for an expanded observing programme is to understand the detailed structure of magnetised plasma in this system, and thereby to provide insight into processes occurring in the FR-I radio galaxy population as a whole. In this paper however, we present preliminary images and results for the southern Cen A lobe, and document its linear polarisation structure at $\sim13\times$ better spatial resolution than has previously been published at any radio frequency (and $\sim40\times$ better for a similar frequency). We note that we have also observed the northern lobe, but these data will be presented elsewhere.\\

We adopt a distance to Cen A of 3.8 Mpc from \citet{HRH2010}. At this distance, 1 arcminute corresponds to a projected linear size of 1.1 kpc. Linear polarisation states are represented by a complex vector $\boldsymbol{P}$ with magnitude $P$, which is related to the Stokes parameters $Q$ and $U$, the polarisation angle $\psi$, the fractional polarisation $p$, and the total emission intensity $I$ as:
 
\begin{equation}
\boldsymbol{P} = Q + iU = pIe^{2i\psi}
\label{eq:ComplexPolVec}
\end{equation}

In clouds of magnetised plasma between a point $L$ and an observer, the Faraday effect can rotate the plane of polarisation of radio waves by an amount equal to  

\begin{equation}
\Delta\psi= \phi\lambda^2
\label{eq:rotation}
\end{equation}

\noindent where $\psi$ is the polarisation angle, $\lambda$ is the observing wavelength, and $\phi$ is the Faraday depth, given by 

\begin{equation}
\text{$\phi$}(L) = 0.812 \int_{L}^{\text{telescope}} n_e\boldsymbol{B}.\text{d}\boldsymbol{s}~\text{rad m}^{-2}
\label{eq:FaradayDepth}
\end{equation}
 
\noindent and $n_e$ [cm$^{-3}$] and $\boldsymbol{B}$ [$\muup$G] are the thermal electron density and magnetic field along the line of sight (LOS) respectively.

The observable $\boldsymbol{P}(\lambda^2)$ is the sum of polarized emission contributions over Faraday depth:

\begin{equation}
\boldsymbol{P}(\lambda^2) = \int_{-\infty}^{\infty} \boldsymbol{F}(\phi) e^{2i\phi\lambda^2} d\phi
\label{eq:SumPol}
\end{equation}
 
$\boldsymbol{F}(\phi)$ --- referred to as the Faraday Dispersion Function --- specifies the distribution of polarised emission over Faraday depth along the LOS. It can be reconstructed using methods such as RM synthesis \citep{Burn1966,BdB2005} and {\sc rmclean} \citep{Heald2009}.

\subsection{The ASKAP radio telescope}\label{sec-askap}

The Australian Square Kilometre Array Pathfinder (ASKAP; \citealp{DeBoer2009,Johnston2007,SB2016}) is a new aperture synthesis array nearing completion in Western Australia. It consists of 36 antennas, a minimum (maximum) baseline of 20m (6.4km), abundant short baselines, a correlated bandwidth of 300 MHz operating in the range 700--1800 MHz, and is currently being ramped up to full capability towards the end of its commissioning phase. ASKAP is designed for wide field imaging: Each antenna is equipped with a Phased Array Feed (PAF; \citealp{HO2008,Hampson2012}) providing an enormous $\sim$ 30 deg$^2$ field of view, which when coupled with ASKAP's superior sensitivity, yields exceptional survey speed and widefield imaging capabilities. Much of its time will therefore be dedicated to survey projects such as the POlarisation Sky Survey of the Universe's Magnetism (POSSUM; Gaensler et al. \emph{in prep.}), in which an expected 1 million polarised sources will be observed in the frequency range 1.1--1.4 GHz, the Evolutionary Map of the Universe (EMU; \citealp{Norris2011}), in which $\sim70$ million radio sources will be observed over the same band in total intensity, and WALLABY \citep{Koribalski2012}, expecting to detect over $5\times10^5$ galaxies in the HI spectral line. Nevertheless, ASKAP is also well positioned to target large individual objects, whose angular sizes make them expensive to observe with traditional synthesis arrays.  

\subsection{These observations}

\begin{figure}
\centering
\includegraphics[width=0.47\textwidth]{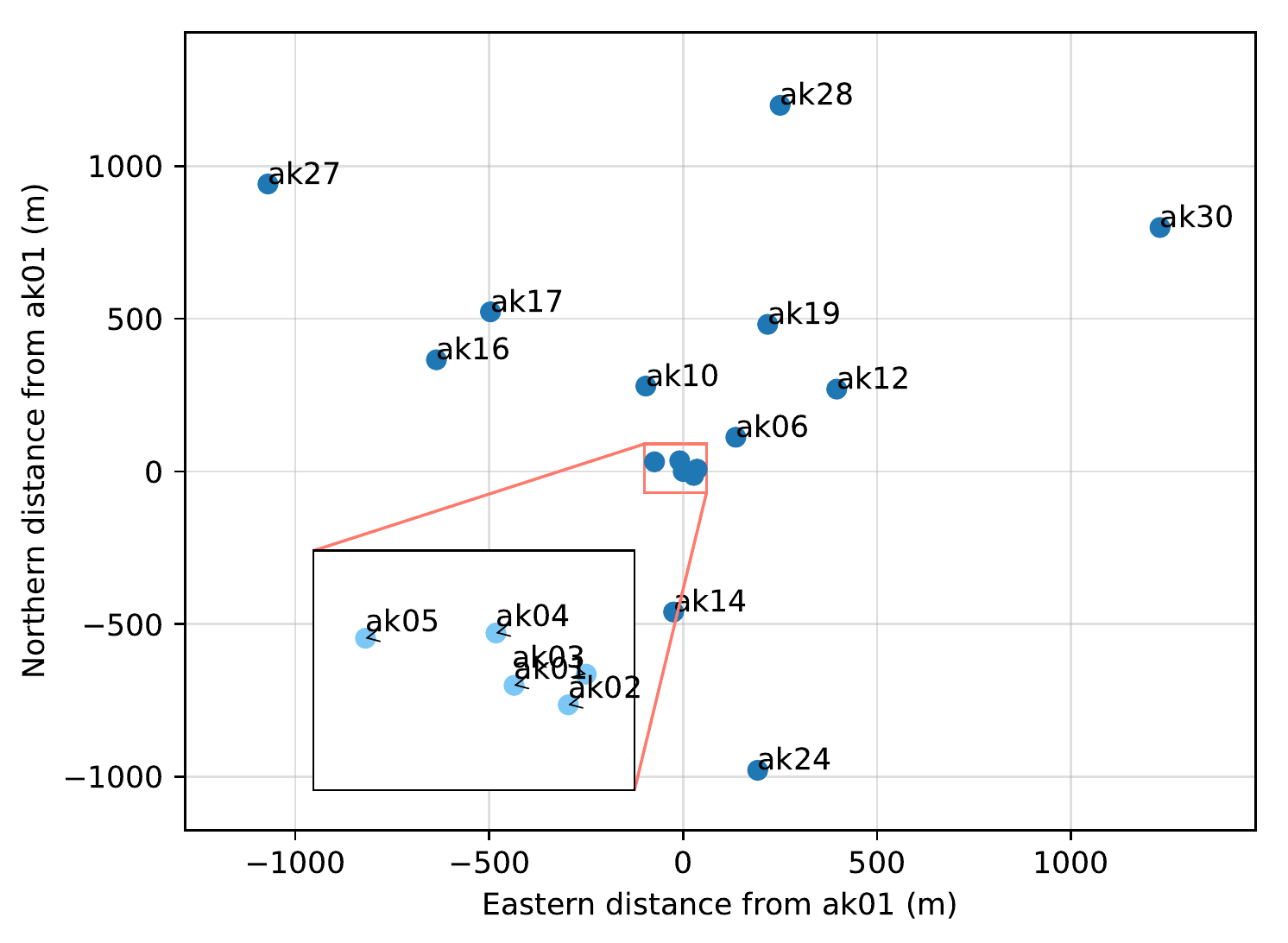}
\caption{Offset ASKAP antenna positions in meters from antenna ak01 for antennas active in our observations. The inset panel zooms in on east-west offsets of -100m to +60m, and north-south offsets of -70m to +90m.}
\label{fig:array_config}
\end{figure}

\begin{figure}
\centering
\includegraphics[width=0.47\textwidth]{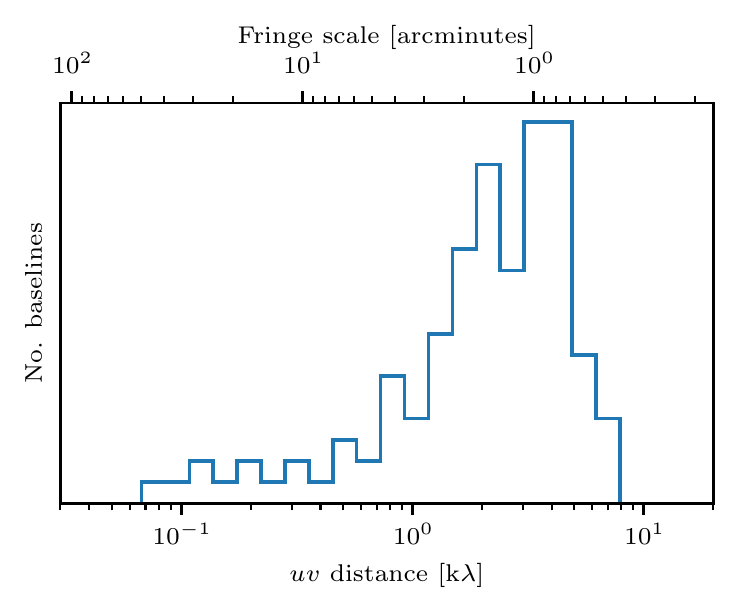}
\caption{Histogram of the number of baselines as a function of $uv$ distance (in k$\lambda$), and the associated angular fringe scale on the sky, for ASKAP-16. The $uv$ distances were calculated at our band centre frequency of 913 MHz.}
\label{fig:bl_histo_logx}
\end{figure}

We exploited ASKAP's unique capabilities to target the southern outer lobe of Cen A as part of ASKAP commissioning activities. We observed the lobe for 12 hours in a single pointing, recording full polarisation data over a 240 MHz bandwidth centred at 913 MHz. At the time, the array consisted of 16 active antennas (ASKAP-16, henceforth), with positions and baselines detailed in Figures \ref{fig:array_config} and \ref{fig:bl_histo_logx}. A summary of the observations is provided in Table \ref{tab:obsdeets}.

\begin{table}
\centering
\caption{ASKAP observing setup}
\label{tab:obsdeets}
\begin{tabular}{ll}
\hline
\hline
Target & Cen. A southern lobe \\
Field centre (J2000) & $13^h25^m00^s$, $-45^d00^m00^s$ \\
Date & 26th January 2018\\
Scheduling block ID & 5036\\
No. telescope pointings & 1 \\
Full-band sensitivity & 85 $\muup$Jy beam$^{-1}$ \\
\quad (measured, robust $=-0.5$) & \\
Recorded polarisations & $XX$, $XY$, $YX$, $YY$ \\
Number of antennas & 16 \\
Antenna diameter & 12 m \\
Longest baseline & 2303 m \\
Shortest baseline & 22.4 m \\
Angular resolution & 26 arcsec $\times33$ arcsec \\
Max. recoverable scale & 30 arcmin\\
Number of beams & 36 paired X and Y pol.\\
Formed beamwidth (900 MHz) & 1.59 deg\\
Total sky coverage & $\sim35$ deg$^2$ \\
Frequency range & 793--1033 MHz \\
Frequency resolution & 1 MHz \\
$\lambda^2$ range & 0.084--0.143 m$^2$ \\
Resolution in $\phi$ space & 59 rad m$^{-2}$ \\
Largest recoverable $\phi$ scale & 37 rad m$^{-2}$ \\
\quad (at 50\% sensitivity) & \\
Max. recoverable |$\phi$| & 7300 rad m$^{-2}$ \\
\hline
\end{tabular}
\end{table}

The experiment reported on here --- concerned solely with imaging the linearly polarised emission from the lobe --- was, in two key respects, enabled by the relative scale-sizes of the polarised versus total emission intensity. Firstly, the characteristic angular emission scale for Stokes $Q$ and $U$ is smaller than 20 arcminutes \citep{Feain2009}. We can therefore reconstruct the polarised emission structure without requiring zero-spacing data. Second, in total intensity, even the most striking small-scale features in the lobes represent only small deviations from the smooth underlying large-scale emission structure --- typically less than a few mJy beam$^{-1}$ \citep{Feain2011}. Thus, the total intensity flux from the outer lobe is almost fully resolved out, and we expect leakage from Stokes $I$ into $Q$, $U$ and $V$ to be negligible \citep{Sault2015,Willis2011}.

\subsubsection{Beam-forming and positioning on the sky}\label{sec:beams}

The ASKAP PAFs \citep{Hampson2012} consist of 188 connected dipole antennas in a chequerboard pattern \citep{HO2008}, divided evenly between orthogonal polarisations. Signals from these dipole elements are linearly combined to form paired (i.e. co-located on the sky) X and Y polarisation beams towards 36 different directions on the sky at the same time. Different beamforming strategies exist to optimise different properties of the formed beams --- we used the `maximum signal-to-noise ratio' algorithm, implemented as described in \citet{McConnell2016}, which is the only PAF beamforming option to date supported for routine ASKAP observations.

\begin{figure*}
\centering
\includegraphics[width=\textwidth]{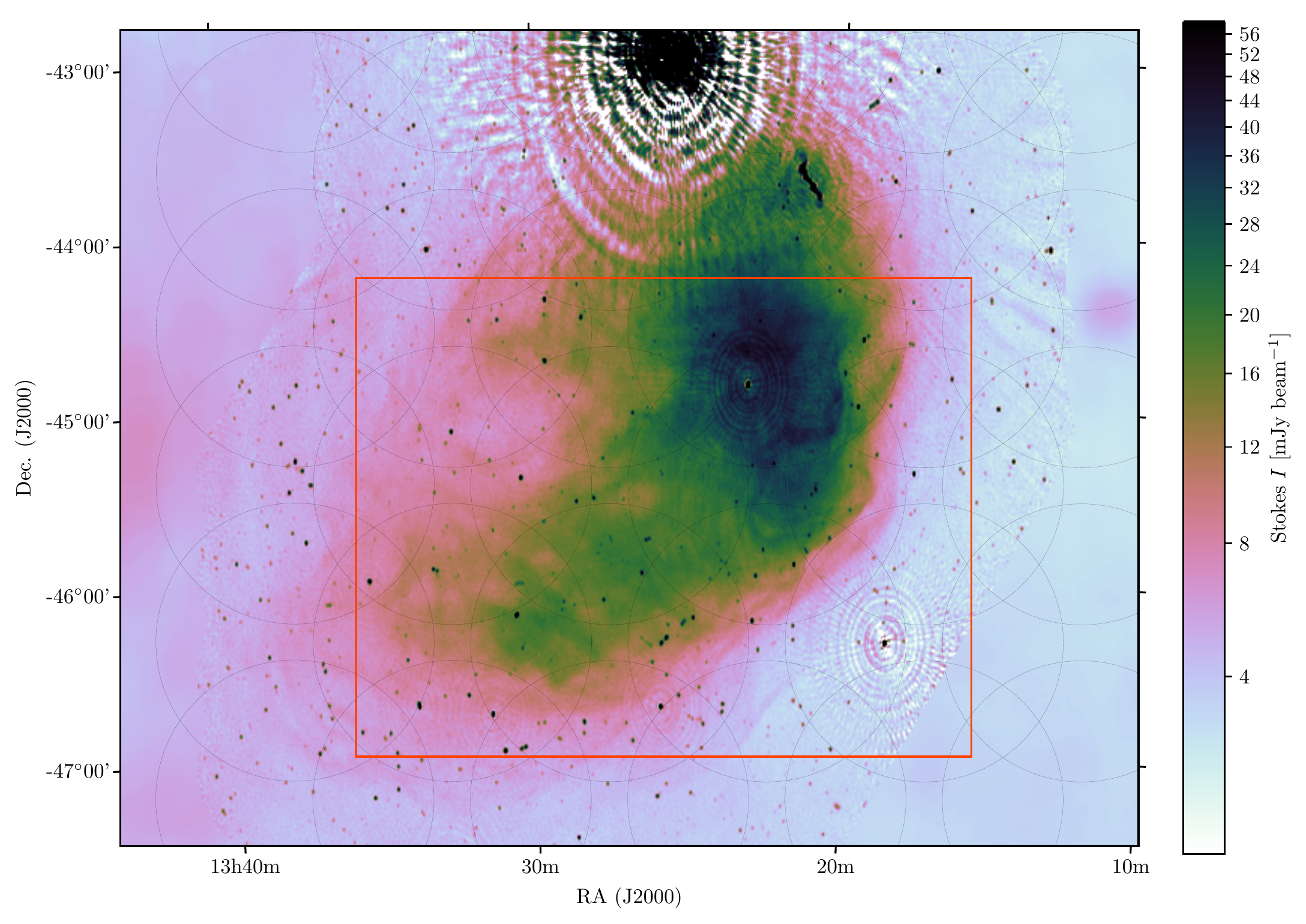}
\caption{Combined Parkes and ATCA Stokes $I$ map of Centaurus A created by \citet{Feain2011}, with $60\times40$ arcsecond spatial resolution and a sensitivity of $200\muup$Jy beam$^{-1}$ (shown with a cubehelix color map; \citealp{Green2011}). The black circles indicate the position and diameter (at 913 MHz) of ASKAP's formed beams for these observations. The red box indicates the region shown in Figs. \ref{fig:p_sl} and \ref{fig:peakFD}.}
\label{fig:cenabeamolay}
\end{figure*}

We tiled the southern Cen A lobe with a square $6\times6$ grid of (paired X and Y) beams, separated by 0.9 degrees along rows and columns (cf. the beamwidth recorded in Table \ref{tab:obsdeets}). The edges of the square footprint were oriented along lines of constant RA and decl., and the roll axis \citep{Forsyth2009,Heywood2014} on each antenna was driven to maintain this orientation on the sky. The beam layout is shown in Figure \ref{fig:cenabeamolay}.

\begin{figure*}
\centering
\includegraphics[width=\textwidth]{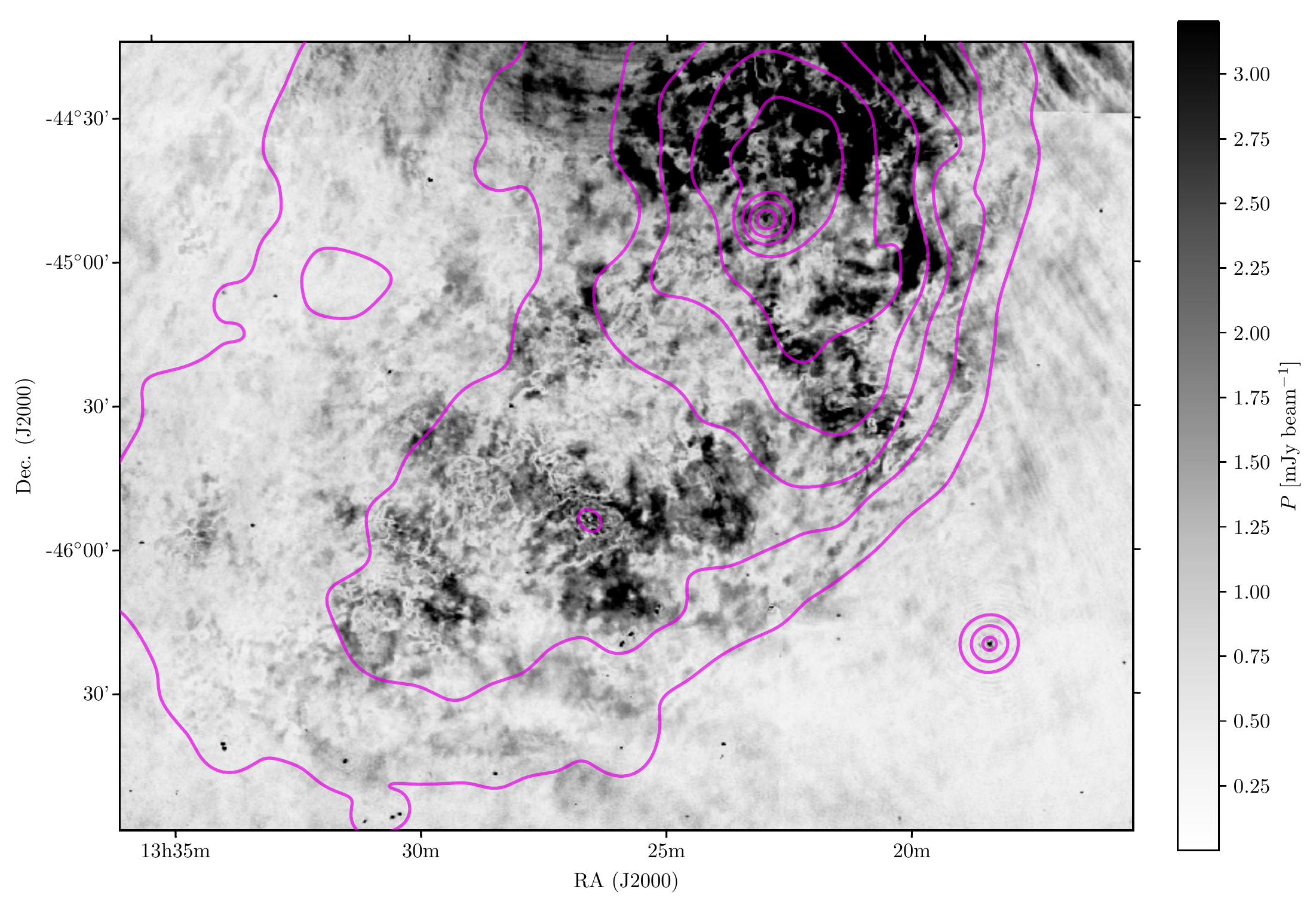}
\caption{Grayscale: ASKAP 913 MHz peak polarised intensity map of the southern lobe of Cen A at 26$\times33$ arcsecond spatial resolution (note the higher spatial resolution compared to Fig. \ref{fig:cenabeamolay}). The measured R.M.S. noise level is 85 $\muup$Jy beam$^{-1}$ (with robust $=-0.5$ weighting). Overlay: Stokes $I$ contours, indicating flux densities of 1.4, 7.5, 13.5, 20, 26, 32, 38, 44, 50, and 56 mJy beam$^{-1}$ at 5 arcminute resolution. The region shown is as indicated in Fig. \ref{fig:cenabeamolay}.}
\label{fig:p_sl}
\end{figure*}

\begin{figure*}
\centering
\includegraphics[width=\textwidth]{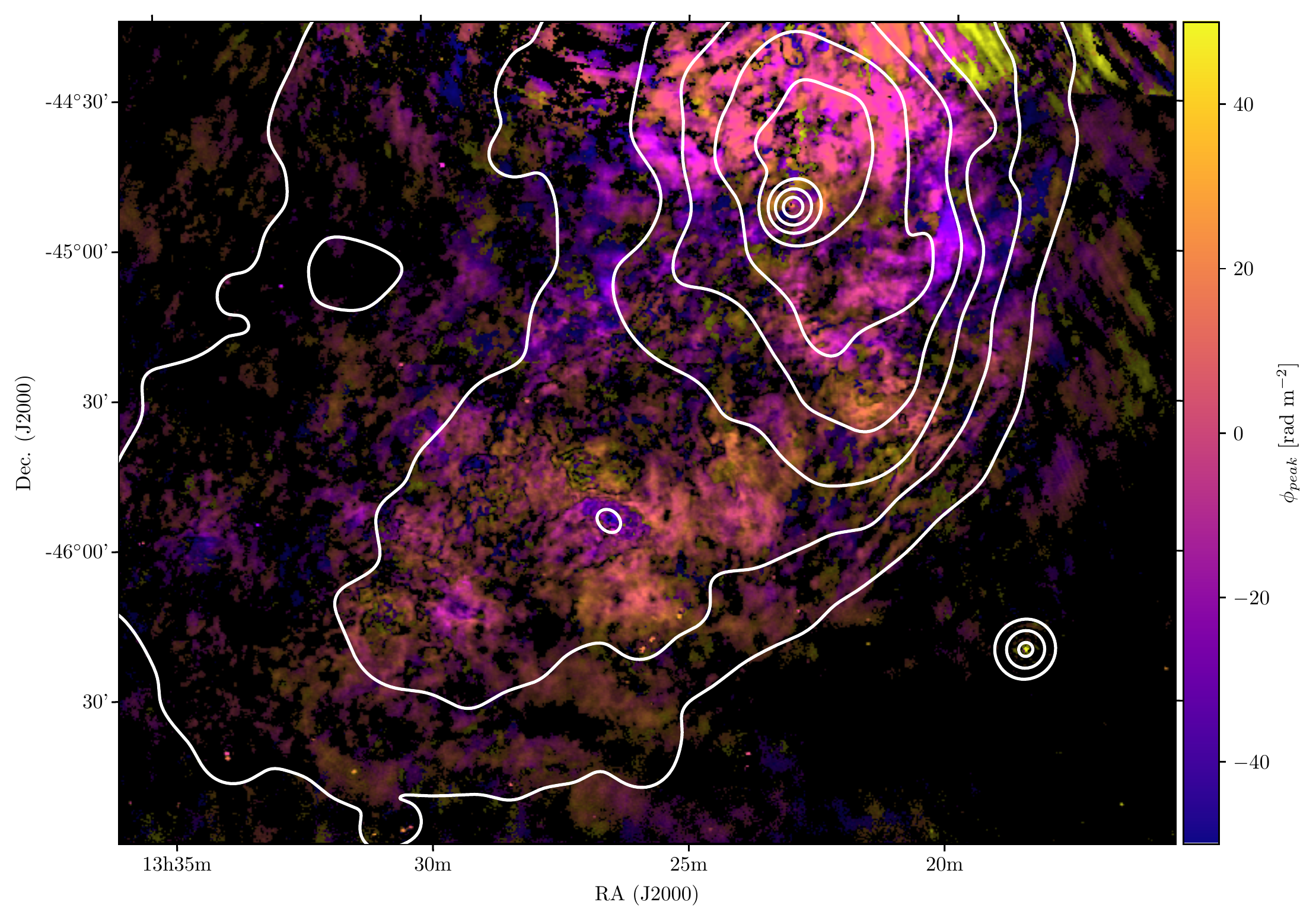}
\caption{Colourscale: ASKAP 913 MHz hue-intensity map of the southern lobe of Cen A. The hue channel traces $\phi_\text{peak}$, while the intensity channel traces $P$. A flat Galactic foreground contribution of $-60$ rad m$^{-2}$ has been subtracted from $\phi_\text{peak}$ \citep{Feain2009}. Overlay: Stokes $I$ contours, indicating flux densities of 1.4, 7.5, 13.5, 20, 26, 32, 38, 44, 50, and 56 mJy beam$^{-1}$ at 5 arcminute resolution. The region shown is as indicated in Fig. \ref{fig:cenabeamolay}.} 
\label{fig:peakFD}
\end{figure*}

\section{Data processing and imaging}

ASKAP's unique data processing requirements are met by ASKAPsoft \citep{McConnell2016} and an associated data reduction pipeline (to be described by Whiting et al. \emph{in prep.}), which are outlined in Section \ref{sec-ascalim}. The polarisation calibration is currently performed with a separate associated pipeline, which we outline in Section \ref{sec-polcal}. 

\subsection{ASKAPsoft processing}\label{sec-ascalim}

We processed our data with version 0.21.0\footnote{https://doi.org/10.25919/5b446f22c46fc} of the ASKAPsoft data reduction pipeline. The pipeline was set to split and flag the calibrator data (three minute tracks on PKS B1934-638 observed sequentially at the different beam centres) and Cen A data, and then to calculate the bandpass corrections. We applied these corrections to the calibrator data \emph{only} (see Section \ref{sec-polcal}), then halted main pipeline processing, executed the polarisation calibration pipeline (Section \ref{sec-polcal}), then continued main pipeline processing, where the data were:

\begin{itemize}
\item averaged to 1 MHz spectral resolution (from the native 18 kHz resolution); 
\item re-flagged;
\item self-calibrated using field point sources in total intensity (phase  only, 2 rounds);
\item imaged in 1 MHz channels through the band to generate spectral cubes in Stokes $I$, $Q$, $U$, and $V$;
\item smoothed to a common frequency-independent resolution, and
\item linearly mosaicked
\end{itemize} 

The W-snapshotting algorithm \citep{Cornwell2012} was used for imaging, with robust $=-0.5$ visibility weighting. Each image channel in the spectral cubes was independently {\sc clean}ed with the Cotton-Schwab algorithm and a single (delta function) scale. We employed a limited form of {\sc clean} masking, where the entire imaged sky area was searched for {\sc clean} components down to a signal-to-noise ratio of 6$\sigma$, after which {\sc clean}ing continued in only those locations where components were detected down to a threshold of 1$\sigma$. We restored the images with a 26$\times$33 arcsecond Gaussian beam (FWHM) with a position angle of 87$^\circ$.

\subsection{Polarisation calibration}\label{sec-polcal}

Calibrating ASKAP polarisation (on the beam axes) entails nulling out the phase difference between the paired X and Y beams, and correcting for polarisation leakage. Most interferometers use injected calibration signals and/or repeated observations of secondary calibrators to solve for these corrections. The former was not available for our observations, whilst the latter is not viable for ASKAP (due to the large number of beams requiring calibration). Different methods are required, which we have implemented in a dedicated polarisation calibration pipeline. This will be described in detail by Anderson et al. (2019; \emph{in prep.}), but the basic procedure is as follows.

During bandpass calibration (Section \ref{sec-ascalim}), the X and Y beam antenna gains are calculated across the array. The XY phase of the reference antenna is unconstrained for this procedure, but must be solved for. This requires at least one X beam and one Y beam in the array to observe a common reference signal. We obtain such a signal by rotating the PAF on the reference antenna by 5 degrees (though, we emphasize, having formed the X and Y beams in identical positions on the sky to the other antennas). This introduces an apparent $\sin$(5$^\circ$) $\approx$~9\% leakage into the X and Y beams from the orthogonal polarisations on the reference antenna, providing a means of measuring their mutual phase. The phase can be nulled out by applying a modified version of the bandpass corrections derived from main pipeline processing (Section \ref{sec-ascalim}). We note that this method is similar to an approach previously used at Westerbork \citep{Weiler1973}. The true leakages are subsequently solved for using the primary calibrator PKS B1934-638, by assuming that it has Stokes $Q=U=V=0$ in our frequency band. This approximation is accurate to $\sim0.1$\% of Stokes $I$ \citep{Schnitzeler2011}. The absolute polarisation angle remains uncalibrated for our observations, but this does not impact the data or analysis presented here, since it is consistent across beams and as a function of frequency. We note that in the future, the ASKAP polarisation calibration procedure will furnish corrections for the absolute polarisation angle, ionospheric Faraday rotation, and the off-axis cross-polarisation response.

\section{Preliminary analysis and results}

\subsection{Linearly polarised intensity}\label{sec-struct}

We applied RM synthesis \citep{Burn1966,BdB2005} to our Stokes $Q$ and $U$ datacubes (Stokes $I$ was not divided out), yielding a datacube containing the reconstructed $\boldsymbol{F}(\phi)$ at each pixel location. From this we extracted the map of the peak polarised intensity (max(|$\boldsymbol{F}(\phi)$|)) shown in Figure \ref{fig:p_sl}.

\subsubsection{Description}\label{sec-pint-desc}

Linearly polarised emission is found throughout the total intensity lobe (Fig. \ref{fig:p_sl}). Its large-scale structure (angular scales above $\sim$tens of arcminutes) near 900 MHz broadly matches that seen in previous lower resolution observations conducted at 4.8 GHz (cf. Fig. 3b of \citealp{Junkes1993}) and 1.4 GHz (cf. Fig. 1 of \citealp{OSullivan2013}). That is, it is distributed in a broken, elongated ring of interconnected patches that runs around the inner edge of the Stokes $I$ emission, but is offset towards the lobe centre by 30 arcminutes or more. The centre of the ring (and lobe) is dominated by a $\sim$30'$\times$60' region of either low polarised emissivity or foreground depolarisation, oriented along the major axis of the lobe viewed in projection. We note that any polarized component that is smooth in Q and U on scales greater than 30 arcminuntes would not be detected in these observations. However, given the extensive Faraday structure detected (Section \ref{sec:fd}), it is unlikely that such a component exists. 

Moving towards the smaller scales that are uniquely probed by our data, and resolving features as small as 20 arcseconds, complex filamentary networks characterised by low polarisation (henceforth, low-$P$) begin to appear (we note that we use the term `filament' and its derivatives in a purely morphological sense). These filaments cut across many of the brighter polarised regions. Conversely, isolated polarised regions also appear and inhabit the larger regions of low polarised emissivity. 

The networks of low-$P$ filaments are of particular interest. Again, we note that similar `depolarisation canals' have been found to affect interferometer images when the maximum recoverable spatial scale is smaller than the size of the emission being imaged. In this case however, the features appear to be largely independent of frequency, suggesting that they are are not simply a manifestation of interferometric spatial filtering. Our planned follow-up observations (Section \ref{sec-ongoing}) will definitively confirm or refute this point. Assuming the depolarised filamentary networks are real, they must trace either complicated structures in lobe's magnetic field, or in magnetised plasma in the foreground. They are most prominent towards the southeast end of the lobe. Some of the individual low-$P$ filaments that comprise them are resolved while others are not. Some appear to transition between these states along their length, or become so tightly packed that they appear to merge. Both instances are best seen at locations where larger ($\sim30$') regions of high and low polarised emissivity abut one another. The larger low-$P$ regions might therefore result from the superposition of many low-$P$ filaments, or their broadening. 

\subsubsection{Structure function}\label{sec-pint-stfunc}

\begin{figure*}
\centering
\includegraphics[width=\textwidth]{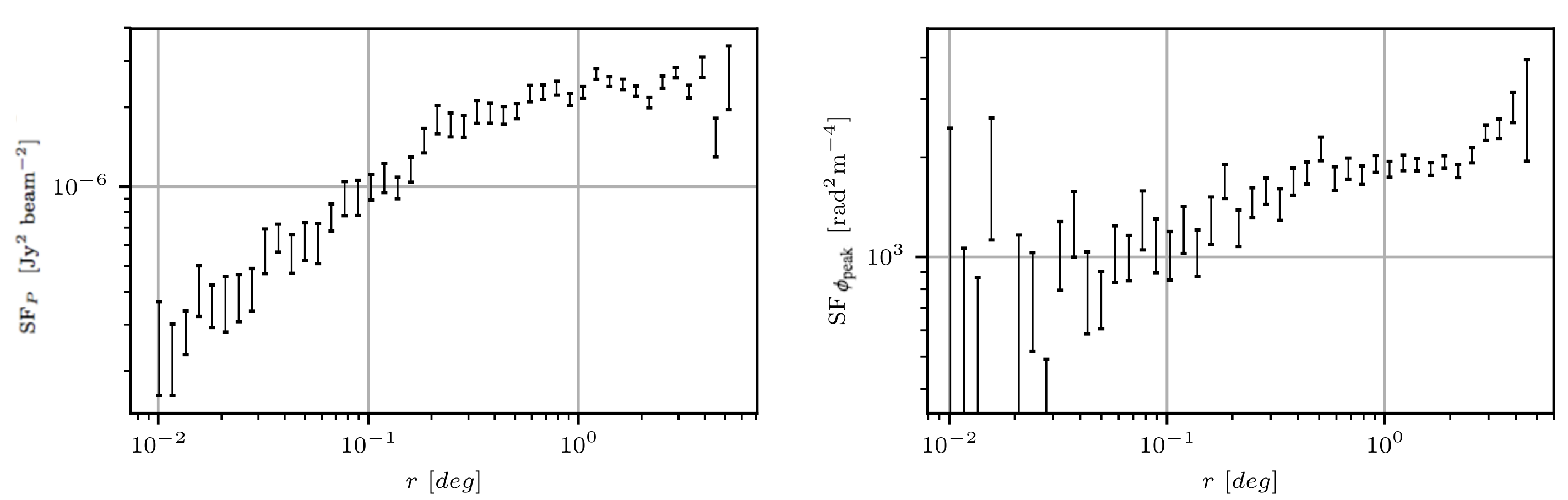}
\caption{Structure function for $P$ (left panel) and $\phi_\text{peak}$ (right panel).}
\label{fig:stfuncs}
\end{figure*}

The degree of scale-dependent structure in $P$ can be parameterised with the second-order, one-dimensional structure function of $P$ (e.g. \citealp{Haverkorn2006}):

\begin{equation}
\text{SF}_{P} = 2\sigma_{P}^2(\theta) = \langle [ P(\theta) - P(\theta+\delta\theta)]^2 \rangle 
\label{eq:strucfuncP}
\end{equation}
 
\noindent
where the effect of measurement uncertainty was removed following Appendix A of \citet{Haverkorn2004}. The result is shown in Figure \ref{fig:stfuncs} (left panel). Above one degree, $\text{SF}_{P}$ shows some scale-dependent structure, but is significantly affected by sampling variance. Below one degree, $\text{SF}_{P}$ is well-characterised by a power law decrease of slope $0.3\pm0.2$ down to $\sim10$ arcminutes (11 kpc), where a break is evident, after which $\text{SF}_{P}$ decreases more rapidly, with a slope $0.7\pm0.1$ down to scales of $\sim30$ arcseconds ($\sim550$ pc). There is no indication that the inner scale of spatial structure in $P$ is being approached at these scales.

\subsection{Faraday depth}\label{sec:fd}

We extracted a map of the peak Faraday depth ($\phi_\text{peak}$) as argmax(|$\boldsymbol{F}(\phi)$|). From \citet{Feain2009}, the average Galactic foreground RM contribution at this location is approximately $-60$ rad m$^{-2}$. We subtracted this value from our data, while acknowledging that the Galactic Faraday screen may vary on smaller scales than this (e.g. \citealp{Leahy1987,Haverkorn2004,Anderson2015,OSullivan2013}). We present the result as a hue-intensity map in Figure \ref{fig:peakFD}, where the hue traces $\phi_\text{peak}$, and the intensity traces $P$. The large-scale structure of $\phi_\text{peak}$ (angular scales above $\sim$tens of arcminutes) broadly matches that reported by \citet{OSullivan2013} at 1.4 GHz: Negative (foreground corrected) values of around $-20$ rad m$^{-2}$ are seen in a band that extends along the northeastern ridge of Stokes $I$ emission from $\sim13^h33^m$, $-46^d00^m$ to $\sim13^h27^m$, $-44^d45^m$, before cutting across the lobe from east to west at Decl. $\approx-44^d45^m$, and finally tracking southwards along the western edge of the polarised emission for $\sim30$ arcminutes. The approximately constant line of declination where this band cuts across the lobe corresponds to a marked drop in the polarised intensity (both absolute and fractional; see e.g. Fig. 3b of \citealp{Junkes1993} and Fig. 5 of \citealp{OSullivan2013} respectively) of the lobe when moving southwards, and after which the polarised emission breaks up into the interconnected patches described in Section \ref{sec-pint-desc}. We speculate that this $\phi_\text{peak}$ discontinuity may mark the boundary between different bubbles of radio plasma, buoyantly uplifted after successive episodes of AGN activity, as is perhaps seen in the northern lobe complex \citep{Morganti1999} but not in the south (though see \citealp{OSullivan2013}). However, more work is required to confirm that this is not a Galactic foreground effect. 

$\phi_\text{peak}$ shows considerable small-scale structure, including interfaces across which it rapidly changes value, and possibly changes sign. On the whole though, there appears to be little evidence for spatial correspondence between the depolarised filaments and structure in $\phi_\text{peak}$, as is seen for example in Fornax A \citep{Anderson2018}. A partial and notable exception is located in a region centred roughly on $13^h26^m30^s$, $-45^d50^m00^s$ in Fig. \ref{fig:peakFD}, extending approximately 15 arcminutes in each direction from this point. Here we see a remarkable structure, with pronounced negative values of $\phi_\text{peak}$ in a roughly circular region to the south, and pronounced positive values of $\phi_\text{peak}$ in an apparently mirrored circular region to the north. Emerging from between these two circular regions and extending to the east and west, we observe two cone-shaped regions characterised by intermediate values of $\phi_\text{peak}$. There are clear interfaces in $\phi_\text{peak}$ between these different regions, and these are spatially coincident with depolarised filaments. Conversely, numerous depolarised filaments inside these regions are evidently not associated with such interfaces.  

Similar to our analysis in Section \ref{sec-pint-stfunc}, we calculated the second-order, one-dimensional structure function of $\phi_\text{peak}$ for the lobe. The result is presented in Figure \ref{fig:stfuncs} (right panel). We observe a power law increase in $\text{SF}_{\phi_\text{peak}}$ from 30 arcseconds up to $\sim0.4$ degrees. Here we see a break, and $\text{SF}_\phi$ becomes flat, suggesting that the outer scale of structure in $\phi_\text{peak}$ has been reached. This is similar to the results obtained by \citet{Feain2009} through a structure function analysis of background point sources. However, while the outer scale is clearly reached at 0.3$^\circ$ in Feain et al.'s data, the outer scale in Figure \ref{fig:stfuncs} is perhaps larger by 0.1--0.2 degrees. If this difference between the respective outer scales can be conclusively established, it might represent direct evidence of Faraday active material being dispersed through the lobe volume. We will pursue this in future work. At two degrees, $\text{SF}_{\phi_\text{peak}}$ begins to rise strongly up to the scale size of the lobe itself. This seems to be due to the large scale structure of $\phi_\text{peak}$ around the periphery of the lobe, as discussed earlier.

\section{Discussion and ongoing work}\label{sec-ongoing}

Our observations show that ASKAP already enjoys excellent polarisation imaging capabilities, even during its commissioning phase. Numerous important and relevant upgrades to the instrument have occurred since the observations presented here, and will continue to be implemented in the short term (see below). Thus, the quality of polarimetric imaging will continue to improve, and usher in an exciting new era in widefield polarisation science.   

Our new Cen A data reveal complicated linear polarisation structure in the southern lobe --- far more complicated than is seen in total intensity (cf. Fig. \ref{fig:cenabeamolay}) --- down to the limit of our observing resolution ($\sim$480 parsecs). These features appear to be mostly frequency-independent, and are therefore probably not caused by missing short interferometer spacings. Our planned observations (see below) will definitively confirm or refute this proposition.

Our structure functions for both $P$ and $\phi_\text{peak}$ show no sign that our observations are close to probing the turbulent inner scale of magnetised gas along the line of sight. Networks of depolarised filaments are observed which indicate the presence of complex magnetic field structure along the line of sight, but only become fully apparent at sub-arcminute scales, with almost 500 resolution elements across the source. Establishing the relative contribution of local (to the lobe) vs. foreground effects in giving rise to these structures in $P$ and $\phi_\text{peak}$ will require more work and further observations. However, it is clear that for this object in particular, and perhaps FR-I-type objects in general, exceptional resolution, frequency coverage, and advanced analysis techniques will be required to help disentangle the structure and location of the magnetised plasmas apparent in these systems. We are currently exploring the use of analysis techniques like the polarisation gradient (\citealp{Gaensler2011}; see also \citealp{Herron2018a}), and related statistical techniques, to attempt to differentiate between Faraday rotation occurring predominantly inside versus outside the lobe \citep{Sun2014,Herron2018b}. The polarisation gradient maps in particular reveal myriad interesting structures, including possible indications of strong shocks occurring in the lobes. These results will be presented elsewhere.

Past results show that the northern and southern outer lobes differ considerably in their radio morphology, including their polarised emission and Faraday rotation structure (e.g. \citealp{Feain2009,OSullivan2013}). Thus, in the near future, we plan to observe the entirety of Cen A with ASKAP-36 (the full ASKAP array) and the new Ultra Wideband Low (UWL) receiver on the Parkes telescope \citep{Manchester2015}. The combination will allow for full Stokes imaging of Cen A at close to 6" spatial resolution (110 pc) at 1700 MHz, at a full-band sensitivity of $\sim$tens of $\muup$Jy beam$^{-1}$. The potential for high-fidelity imaging is excellent: The quality of raw ASKAP data has improved substantially in many respects since the observations reported here. The additional use of techniques like shape-constrained beamforming, joint imaging and deconvolution, A-projection, and direction-dependent gain calibration are planned, and will improve image quality and polarisation purity further still. The upgraded Murchison Widefield Array \citep{Wayth2018} will provide valuable ancillary data in the form of greatly expanded $\lambda^2$ coverage, and thus a huge lever arm for detailed rotation measure and depolarisation analysis (e.g. \citealp{OSullivan2012,Anderson2015, Anderson2016,Lenc2017,OSullivan2017,OSullivan2018,Riseley2018}). At higher frequencies, instruments such as the Australia Telescope Compact Array and MeerKAT can conduct detailed multi-wavelength broadband observations of small-scale structures in the lobe to better constrain their nature, such as the low-$P$ filaments. In the longer term, the Square Kilometre Array (SKA) will expand on all of these capabilities, and allow us to study Cen A's more distant analogs in a similar manner (e.g. \citealp{Gaensler2015}).

\begin{acknowledgements}

The Australian SKA Pathfinder is part of the Australia Telescope National Facility which is managed by CSIRO. Operation of ASKAP is funded by the Australian Government with support from the National Collaborative Research Infrastructure Strategy. Establishment of the Murchison Radio-astronomy Observatory was funded by the Australian Government and the Government of Western Australia. ASKAP uses advanced supercomputing resources at the Pawsey Supercomputing Centre. We acknowledge the Wajarri Yamatji people as the traditional owners of the Observatory site. The POSSUM project has been made possible through funding from the Australian Research Council, the Natural Sciences and Engineering Research Council of Canada, the Canada Research Chairs Program, and the Canada Foundation for Innovation. SPO acknowledges financial support from the Deutsche Forschungsgemeinschaft (DFG) under grant BR2026/23. Partial support for LR is provided through U.S. NSF Grant 1714205 to the University of Minnesota.

\end{acknowledgements}

\bibliographystyle{pasa-mnras}
\bibliography{bibliography}

\end{document}